\documentclass[12pt,preprint]{aastex}
\shorttitle{Galaxy Spins in cDE Models}
\begin{document}
\title{Massive Neutrinos Promote the Size Growth of Early-Type Galaxies}
\author{Hyunmi Song and Jounghun Lee}
\affil{Astronomy program, FPRD, Department of Physics and Astronomy,
Seoul National University, Seoul 151-747, Republic of Korea \\
\email{hmsong@astro.snu.ac.kr, jounghun@astro.snu.ac.kr}}
\begin{abstract} 
The effect of massive neutrinos on the evolution of the early type galaxies 
in size and stellar mass  is explored by tracing the merging history of galaxy 
progenitors with the help of the robust semi-analytic prescriptions. 
We show that as the presence of massive neutrinos plays a role of enhancing 
the mean merger rate per halo as well as the merger driven increment in halo mass, 
the high-$z$  progenitors of a massive descendant galactic halo evolve more 
rapidly in mass-normalized size for a $\Lambda$MDM ($\Lambda$  Cold Dark Matter 
+ massive neutrinos) model than for the $\Lambda$CDM ($\Lambda$ Cold Dark Matter) case.  
We provide a physical reason for why the halo mass growth rate and the merger rate are higher in a 
$\Lambda$MDM cosmology and conclude that if the presence and role of massive neutrinos are 
properly taken into account  it may explain away the anomalous compactness of the high-$z$ 
massive ETGs compared with the local giant ellipticals with similar stellar masses.
\end{abstract}

\section{INTRODUCTION}
\label{intro}

It has been recently discovered that the quiescent early type galaxies (ETGs) 
at high redshifts ($z\ge 1$) are smaller in size ($R_{e}$) by a factor of 
($3$-$5$) compared with the local giant ellipticals of similar stellar masses 
($M_{\star}$) \citep{daddi-etal05,trujillo-etal06,vandokkum-etal08}, which 
implies that the high-$z$ massive ETGs must have evolved very strongly in 
size but not in stellar masses. 
In the beginning it was suspected that the stellar masses of 
the high-$z$ ETGs from the photometric data should be overestimated. But  
the subsequent spectroscopic estimates of their dynamical masses 
disproved this suspicion, confirming the rapid size growth of the 
high-$z$ massive ETGs \citep{van-de-san11}. 
It has been a puzzling mystery what mechanism drove such strong size evolution 
of the ETGs without triggering their star formation activity. 

\citet{fan-etal08,fan-etal10} proposed a scenario that the compact sizes of 
the high-$z$ massive ETGs should be attributed to the AGN feedback effect of 
blowing off the baryonic gas out of the halo potential wells. The merit of this scenario 
was that it can explain naturally why the compact ETGs are observed at 
high redshifts $z> 1$ when the quasars were most energetic.  
However, \citet{ragone-figueroa-granato11} cast a doubt on this 
scenario based on the argument that it is difficult to understand in this 
scenario why the most compact high-$z$ ETGs are the 
quiescent ones possessing old stellar populations.

An alternative popular scenario was that the progenitors of the present local 
giant ellipticals underwent very frequent dry mergers through which they grew 
rapidly in size but only mildly in stellar mass 
\citep{hopkins-etal09,khochfar-silk06,nipoti-etal03,nipoti-etal09a,
nipoti-etal09b,oser-etal12,vanderwel-etal09}. 
At first glance this scenario seems quite plausible, fitting well into the 
standard theory of hierarchical structure formation based on the $\Lambda$CDM 
($\Lambda$ + cold dark matter) cosmology. 
But, recently \citet{nipoti-etal12} have revealed that 
this scenario in fact challenges the standard model rather than supporting 
it since even the maximum merger-driven size growth predicted by the $\Lambda$CDM 
cosmology is not fast enough to catch up with the observational trend 
\citep[see also][]{cimatti-etal12}, unless the key cosmological parameters 
deviate substantially from the WMAP7 values \citep{wmap7}.

We note, however, that when \citet{nipoti-etal12} calculated the maximum merger-driven 
size growth consistent with $\Lambda$CDM cosmology merger rate, they missed one crucial 
point that the real structure formation in the Universe may not proceed in 
a perfectly hierarchical way due to the presence of small amount of hot dark 
matter components --- massive neutrinos. 
Since it was found and confirmed that the neutrinos are not massless
\citep{ahmad-etal01,arnett-rosner87,cleveland-etal98,fogli-etal06,
fukuda-etal98,maltoni-etal04}, plenty of theoretical works have been 
devoted to investigating what effect the massive neutrinos would have on 
the formation and evolution of the cosmic structures and how significant
the effect would be \citep[for a recent review, see][]{LP06}. 

It is now understood that on the time-dependent scales where the massive 
neutrinos become non-relativistic, the high velocity dispersion of massive 
neutrinos would damp out the dark matter (DM) halo potential wells, resulting in the 
suppression of the formation of DM halos \citep{abazajian-etal05,
agarwal-feldman11,arhipova-etal02,bird-etal12,ichiki-takada12,
lesgourgues-etal09,mantz-etal10,marulli-etal11,saito-etal08,saito-etal09}. 
The recent works of \citet{SL11,SL12} also suggested that the effect of massive 
neutrinos should also bring about non-negligible change in the mean merging 
rate of DM halos. Assuming a strong dependence of the size growth of the observed 
ETGs on the merging rate of their underlying DM halos, it is logically expected 
that the ETG compactness would be also affected by the presence 
of massive neutrinos. 

Here we claim that the presence of massive neutrinos ($\nu$) could 
significantly promote the size growth of the massive ETGs, explaining away the 
observed anomalous compactness of massive ETGs at high redshifts. 
To prove our claim, we trace backward the progenitor evolution of a 
massive descendant galactic halo, just as \citet{nipoti-etal12} did, but taking the effect of 
massive neutrinos into account. The organization of this Paper is as follows.
In section \ref{sec:review} we provide a brief review of the model constructed 
by \citet{nipoti-etal12} for the size and stellar mass evolution of the ETGs.
In section \ref{sec:analysis} we present a new model for the size 
and stellar mass evolution of the ETGs for a $\Lambda$MDM ($\Lambda$ Mixed Dark 
Matter which represents CDM $+$ $\nu$) 
cosmology and explain how the observed compactness of the massive high-$z$ 
ETGs can be explained away by taking into account the presence of massive neutrinos.
In section \ref{sec:con} we summarize our result and discuss its caveat and 
cosmological implication as well.

\section{MODEL OF THE ETG PROGENITOR EVOLUTION: A REVIEW}
\label{sec:review}

\citet[][hereafter N12]{nipoti-etal12} envisaged a simple picture where a
dissipationless dry merger (which maximizes the size growth of the 
progenitor galaxies) occurs between a main galaxy of mass 
$M_{\rm  main}(z)$ and a satellite galaxy of mass $M_{\rm sat}(z)$ to
form a merged galaxy of mass $M_{h}(z)$ at redshift $z$, which represents 
a progenitor of a descendant galaxy of mass $M_{h}(z_{0})$ observed at a given 
redshift $z_{0} < z$. Then, they modeled the merger-driven evolution
of the size and stellar mass of the progenitor galaxy as  
\begin{eqnarray}
\label{eqn:d2Msdzdxi}
\frac{d^2M_\star}{dzd\xi}
&=&\frac{dM_\star}{dM_h}\left(\frac{d^2M_h}{dzd\xi}\right)_{\rm merg},\\
\label{eqn:d2Redzdxi}
\frac{d^2{\rm ln}R_e}{dzd\xi}&=&
\frac{d\ln R_e}{d\ln M_\star}
\frac{1}{M_\star}\frac{dM_\star}{dM_h}
\left(\frac{d^2M_h}{dzd\xi}\right)_{\rm merg}, 
\end{eqnarray}
where $\xi\equiv M_{\rm sat}/M_{\rm main}$ and the subscript "merg" in the 
right-hand side means the change caused by the halo merging events not by the smooth 
accretion of matter. Here the progenitor size $R_{e}$ represents the effective radius that 
encloses half the progenitor stellar mass. 

To evaluate Equations (\ref{eqn:d2Msdzdxi})-(\ref{eqn:d2Redzdxi}), 
N12 prescribed the three ingredients: 
(i) the evolution of halo mass $d^{2}M_{h}/(dzd\xi)$; (ii) the stellar-to-halo mass 
relation $dM_{\star}/dM_{h}$; (iii) the stellar mass to size relation 
$dM_{\star}/dR_{e}$. The first ingredient $d^{2}M_{h}/(dzd\xi)$  was prescribed as  
\begin{equation}
\label{eqn:dMhdzdxi_merg}
\left(\frac{d^2M_h}{dzd\xi}\right)_{\rm merg}
=\frac{\xi M_h(z)}{1+\xi}\frac{d^2N_{\rm merg}}{dzd\xi}(M_h,z,\xi)\, ,
\end{equation}
where $\xi M_h(z)/(1+\xi)$ is the amount of mass that a progenitor 
gathers during a merging event (i.e., the merger-driven increment in halo mass)
that occurs in a redshift interval of $[z,\ z+dz]$ with satellite-to-main mass ratio $\xi$, 
while $d^2N_{\rm merg}/(dzd\xi)$ represents the mean number of such merging
events per a merged halo of mass $M_h(z)$ (i.e., mean merger rate per
halo). From here on, we denote the mean merger rate per halo by 
$\tilde{B}(M_{h},z,\xi)$. 

The tricky part in evaluating Equation (\ref{eqn:dMhdzdxi_merg}) is
the quantity $M_h(z)$ in the right hand side, which represents the
halo accretion history, i.e., the total halo mass of a progenitor at 
$z$, which grows not only through merging between the main and the
satellite galaxies but also through diffusive accretion of matter. 
Thus, the functional form of $M_{h}(z)$ has to be first determined 
for the evaluation of Equation (\ref{eqn:dMhdzdxi_merg}) rather than
regarding it as a integration variable.  
N12 used the fitting formula for $\tilde{B}$ and $M_{h}(z)$ derived 
numerically by \citet[][hereafter FM08]{FM08} from the Millennium
simulations for a $\Lambda$CDM cosmology \citep{millennium05}. 

As for the prescription of the second ingredient $dM_{\star}/dM_{h}$, 
admitting that no standard model has yet to be established for the 
stellar-to-halo mass relation (SHMR), N12 considered three different SHMRs 
provided by \citet[][B10 hereafter]{B10}, \citet[][L12 hereafter]{L12} and 
\citet[][W11 hereafter]{wake-etal11}, respectively. 
These three SHMRs were all obtained by matching various physical and statistical 
properties of the galactic halos from the high-resolution simulations to those of the 
observed galaxies. The difference among the three SHMRs comes from the different redshift 
ranges to which the matching between the numerical and the observational results 
was applied. In the current work, we consider only the first two SHMRs (B10 and L12), 
excluding the last one (W11) which is valid only for relatively narrow redshift range.

The B10 and the L12 SHMRs have the following same functional form in the
redshift range of $1\le z\le 2$:
\begin{equation}
\label{eqn:SHMR}
{\rm log}M_h(M_\star)={\rm log}M_1+q{\rm log}
\left(\frac{M_\star}{M_{\star,0}}\right)
+\frac{\left(M_{\star}/M_{\star,0}\right)^p}
{1+\left(M_{\star}/M_{\star,0}\right)^{-\gamma}}
-\frac{1}{2}.
\end{equation}
Here the values of the five model parameters, $M_{1}$, $M_{\star,0}$,
$q$, $p$, and $\gamma$, vary with the scale factor $a$ as
\begin{eqnarray}
\label{eqn:para1}
{\rm log}M_1(a)&=&M_{1,0}+M_{1,a}(a-1), \\
\label{eqn:para2}
{\rm log}M_{\star,0}(a)&=&M_{\star,0,0}+M_{\star,0,a}(a-1), \\
\label{eqn:para3}
q(a)&=&q_0+q_a(a-1), \\
\label{eqn:para4}
p(a)&=&p_0+p_a(a-1), \\
\label{eqn:para5}
\gamma(a)&=&\gamma_0+\gamma_a(a-1).
\end{eqnarray}
At $z\ge 1$, $M_{\star,0}$ becomes time dependent as
\begin{equation}
\label{eqn:para6}
{\rm log}M_{\star,0}(a)=M_{\star,0,0}+M_{\star,0,a}(a-1)+M_{\star,0,a^2}(a-0.5)^2.
\end{equation}
The best-fit values of these model parameters are provided in Table 2 in B10 
and Table 5 in L12, respectively. Although the original best-fit parameters 
of the L12 SHMR is valid only at $z\le 1$, \citet{nipoti-etal12} extrapolated 
the validity of the L12 SHMR to higher redshifts of $z>1$ and 
redetermined the best-fit parameters. 

The third ingredient, the size-to-stellar mass relation $dR_{e}/dM_{\star}$, 
was prescribed as 
\begin{equation}
\label{eqn:dRedMs}
\frac{d\ln R_e}{d\ln M_\star}=\left[2-\frac{{\rm ln}(1+\xi^{1.4})}
{{\rm ln}(1+\xi)}\right]\, .
\end{equation}
which is a first-order approximation obtained under the following simplified assumptions 
on the merging process \citep{naab-etal09,oser-etal12}: 
the merging halos have spheroidal shapes; the satellites follow the
parabolic orbits; there is no energy loss during the dissipationless
dry merging events.  Plugging the prescribed three ingredients into Equations 
(\ref{eqn:d2Msdzdxi})-(\ref{eqn:d2Redzdxi}) and integrating them 
over $\xi$ and $z$, N12 have finally made the quantitative predictions of 
the standard $\Lambda$CDM cosmology for the merger driven evolution of the 
size and stellar mass of the progenitors $R_{e}(z)$ and $M_{\star}(z)$ 
of a given descendant halo at $z_{0}<z$. Plotting the locations of the high-$z$ 
progenitors at $z\ge 2$ in the $R_{e}(z)$-$M_{\star}(z)$ plane and comparing 
them with the local relations (see Figure 10 in N12), they demonstrated clearly that the 
sizes of the high-$z$ progenitors in $\Lambda$CDM universe 
is much larger than the observed ones. 

Estimating the uncertainties in $R_{e}(z)$ and $M_{\star}(z)$ due to the 
simplified assumptions that they made to prescribe $dR_{e}/dM_{\star}$ 
and $dM_{h}/dM_{\star}$, N12 showed that the most severe uncertainties come 
from the SHMRs and that the disagreement between the $\Lambda$CDM prediction 
and the observational result is robust against the SHMR uncertainties. 
In other words, it was confirmed that although several simplified assumptions 
were made in their determination of the size and stellar mass evolution of 
the progenitor galaxies, the large discrepancy between the $\Lambda$CDM prediction 
and the observational result is not due to the inaccurate modeling of 
$dR_{e}/dM_{\star}$ and $dM_{h}/dM_{\star}$ .

In the follow-up work, \citet{cimatti-etal12}  quantified the compactness of a progenitor galaxy 
as the mass-normalized size of $R_{e}M^{-0.55}_{11}$ with $M_{11}\equiv M_{\star}/(10^{11}M_{\odot})$, 
and  modeled the evolution of the progenitor compactness as a power-law scaling with redshift:  
$R_{e}M^{-0.55}_{\star}\propto (1+z)^{\beta}$, given that the effective sizes of the observed galaxies 
scale as apower-law of their stellar masses, $R_{e}\propto M^{0.55}_{\star}$ 
\citep[see also][and references therein]{newman-etal12}. 
\citet{cimatti-etal12} determined the best-fit slope to be $\beta\approx -0.6$ for the ETGs in the 
redshift range of $0\le z\le 2$ and to be of $\beta\approx -1$ for the ETGs at $0<z<2.6$, 
confirming the rapid evolution of the ETG compactness at high redshifts.  

\section{EFFECT OF MASSIVE NEUTRINOS ON THE ETG PROGENITOR EVOLUTION}
\label{sec:analysis}

In this section we investigate how the presence of massive neutrinos affects  
the compactness evolution of the massive ETGs, applying the N12 model to the 
case of a $\Lambda$MDM cosmology.  
Basically, we compute Equation (\ref{eqn:d2Msdzdxi})-(\ref{eqn:d2Redzdxi}) 
in a similar manner but with a modified prescription for the first 
ingredient, $d^2M_{h}/dzd\xi$, by incorporating the presence of massive neutrinos 
into the picture. 
First of all, we would like to determine the mean merger rate per halo $\tilde{B}$ 
and the halo accretion history $M_{h}(z)$ for the $\Lambda$MDM case.  
Of course, the most accurate way to determine these quantities should be 
to run repeatedly high-resolution N-body simulations for $\Lambda$MDM models 
with various different  values of the neutrino mass $\sum m_{\nu}$ and 
then to determine the empirical fitting formula for $\tilde{B}(z)$ and 
$M_{h}(z)$ as a function of $\sum m_{\nu}$ by analyzing the N-body results. 
Since such simulations are not available at the moment, we utilize the less accurate but 
practical semi-analytic formula for $\tilde{B}$ and $M_{h}$ derived from 
previous works.

\citet{zhang-etal08} analytically derived the following formula for the mean 
merger rate per halo by incorporating the ellipsoidal collapse dynamics into 
the extended Press-Schechter (EPS) theory \citep{PS74,LC93}:  
\begin{equation}
\label{eqn:Be}
\tilde{B}(M_h,z,\xi)=
\tilde{B}_{\rm sph}(M_h,z,\xi) \times		     
A_0{\rm exp}\left(-\frac{A_1^2\tilde{S_i}}{2}\right)
\left\{1+A_2\tilde{S_i}^{3/2}
\left[1+\frac{A_1\tilde{S_i}^{1/2}}{\Gamma(3/2)}\right]\right\}, 
\end{equation}
where $A_0=0.866(1-0.133\nu_0^{-0.615})$, 
$A_1=0.308\nu_0^{-0.115}$, $A_2=0.0373\nu_0^{-0.115}$,
$\nu_0=\omega^2(z)/S(M_h)$, and $\tilde{S_i}=\Delta S_i/S(M_h)$.
Here, $\tilde{B}_{\rm sph}$ represents the original EPS model 
based on the spherical collapse dynamics \citep{LC93}: 
\begin{equation}
\label{eqn:Bs}
\tilde{B}_{\rm sph}(M_h,z,\xi)
=\frac{d\delta_c(z)}{dz}\frac{M_h^2}{(1+\xi)^2M_i}
\frac{dS(M_i)}{dM_i}\frac{1}{\Delta S_i (2\pi\Delta S_i)^{1/2}},
\end{equation}
where $\delta_c(z)=\delta_c/D(z)$, $\delta_c=1.68$, $D(z)$ is the linear 
growth factor, $\Delta S_i\equiv S(M_i)-S(M_h)$, and $S(M)\equiv
\sigma^2(M)$ is the variance of the linear density field smoothed 
on the mass scale of $M$. The quantity $M_i$ in Equation (\ref{eqn:Be}) 
represents the mass of a merging halo which could be either a main 
or a satellite. 
\citet{zhang-etal08} showed that Equation (\ref{eqn:Be}) significantly improves 
its spherical counterpart, Equation (\ref{eqn:Bs}), agreeing much better 
with the N-body results.

We adopt this ellipsoidal EPS model to evaluate the mean merging rate 
per halo $\tilde{B}$ for a $\Lambda$MDM cosmology whose linear density power 
spectrum is characterized by the neutrino mass fraction, 
$f_{\nu}\equiv\Omega_{\nu}/\Omega_{m}$ (where $\Omega_{m}$ and $\Omega_{\nu}$ are 
the matter and the massive neutrino density parameter, respectively).  
Extrapolating the validity of the ellipsoidal EPS model to $\Lambda$MDM cosmology 
may be justified by the results of \citet{marulli-etal11} which showed that the PS-like 
approaches are valid for the evaluation of the halo mass function even 
for  $\Lambda$MDM cosmologies.
For the $\Lambda$MDM linear density power spectrum and growth factor, we 
utilize the analytic formula given by \citet{EH99}. To normalize the 
$\Lambda$MDM linear power spectrum, we make its amplitude 
satisfy $\sigma_{8}=0.8$ for the case of $f_{\nu}=0$ (i.e., without 
massive neutrinos), which is equivalent to the large-scale normalization. 
The other key cosmological parameters are set at the WMAP7 values \citep{wmap7}. 

\begin{figure}
\begin{center}
\plotone{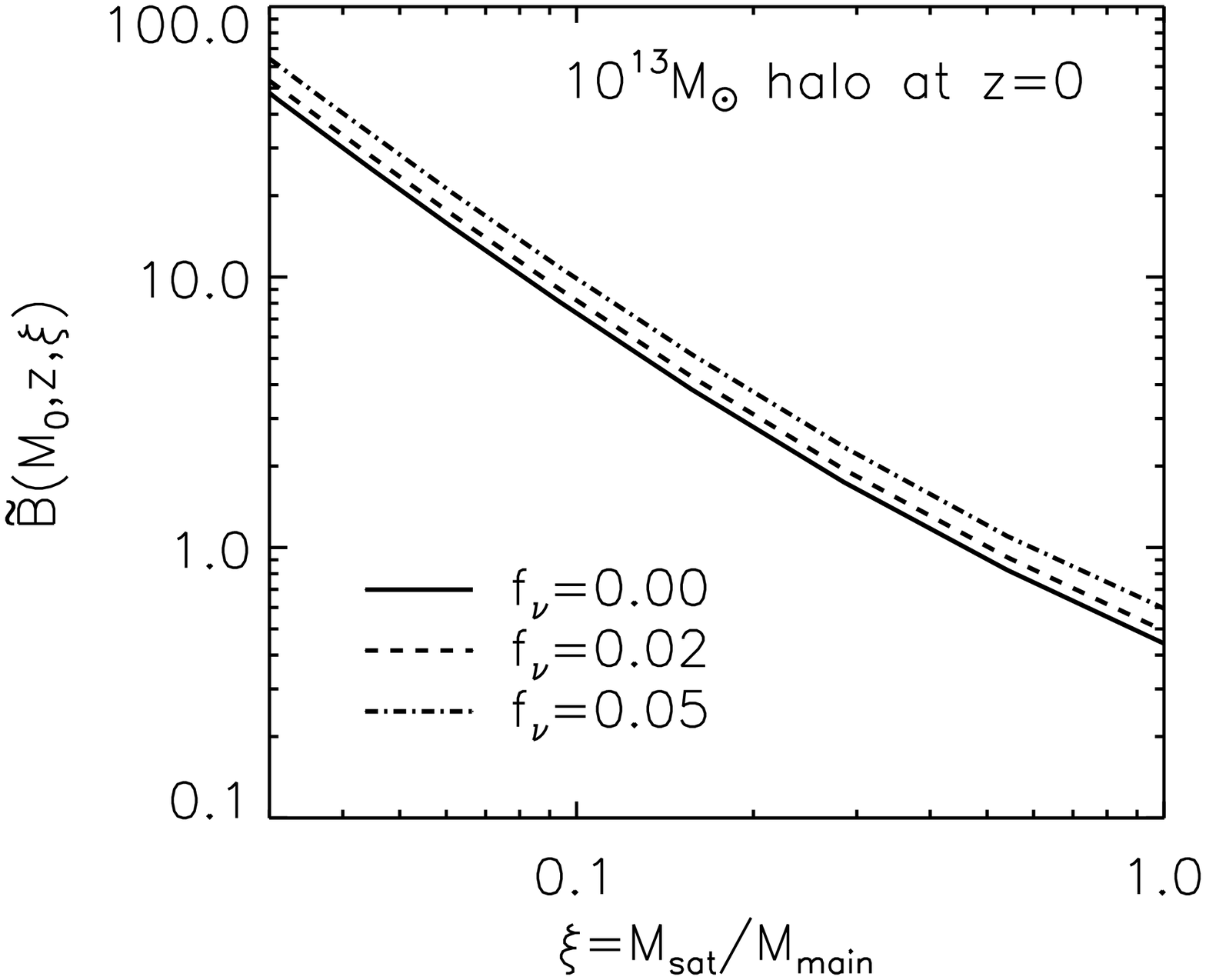}
\caption{Mean merger rate per halo $\tilde{B}$ versus the mass ratio $\xi$ 
for three different cases of the neutrino mass fraction, $f_{\nu}$.}
\label{fig:B}
\end{center}
\end{figure}

Figure \ref{fig:B} shows the mean merger rate per halo as a function of 
$\xi$, for three different cases of $f_{\nu}$ with $M_{h}(0)=10^{13}M_{\odot}$ 
(which corresponds to $M_{\star}(0)\approx 10^{11}h^{-1}M_{\odot}$). 
As can be seen, the larger the neutrino mass fraction is, the higher the mean 
merging rate per halo is. 
For the case of $f_{\nu}=0.05$ (which corresponds to the WMAP7 upper limit 
of $\sum m_{\nu}<0.58\rm{eV}$ with effective number of massive neutrino 
$N_{eff}=4.34$), the overall merger rate per halo increases by a factor 
of $3$ relative to the case of $f_{\nu}=0$.  It is worth mentioning here that 
the redshift $z_{0}$ at which the descendant galactic halo is observed 
is set at the present epoch, i.e., $z_{0}=0$, unlike $z_{0}=1$ in the original 
N12 model, as we are interested in the full progenitor history from the 
high redshifts to the present epoch rather than focusing on explaining 
the high-$z$ phenomena.

As for the mass accretion history for a $\Lambda$MDM universe, we use 
the analytic formula proposed by \citet{zhao-etal09}. Noting the 
existence of a simple expression for the mass accretion rate as a function 
of halo mass, redshift and cosmological parameters, \citet{zhao-etal09} 
developed a theoretical model for the mass accretion history which has
the following universal form:
\begin{equation}
\label{eqn:zhao}
\frac{d{\rm log}\sigma(M)}{d{\rm log}\delta_c(z)}=
\frac{\omega(z,M)-p(z,z_{\rm obs},M_{\rm h}(z_{0}))}{5.85}\, ,
\end{equation}
Here, 
\begin{eqnarray}
\label{eqn:zhao1}
\omega(z,M)&=&\frac{\delta_c(z)}{\sigma(M)10^{d\log\sigma/d\log M}}\, ,\\
\label{eqn:zhao2}
p(z,z_{0},M_{\rm h})&=&p(z_{0},M_{\rm h}(z_{0})) \times
{\rm Max}\left[0,1-\frac{{\rm log}\delta_c(z)-
{\rm log}\delta_c(z_{\rm 0})}{0.272/\omega(z_{\rm 0},M_{\rm h}(z_{0}))}\right],\\
\label{eqn:zhao3}
p(z_{0},M_{\rm h}(z_{0}))&=&
\frac{1}{1+[\omega(z_{0},M_{\rm h}(z_{0}))/4]^6}
\frac{\omega[z_{0},M_{\rm h}(z_{0}])}{2}\, .
\end{eqnarray}
Note that in Equation (\ref{eqn:zhao}), $1/\sigma(M)$ is used as a 
mass-like variable and $1/\delta_c(z)$ is used as a time-like variable. 

\begin{figure}
\begin{center}
\plotone{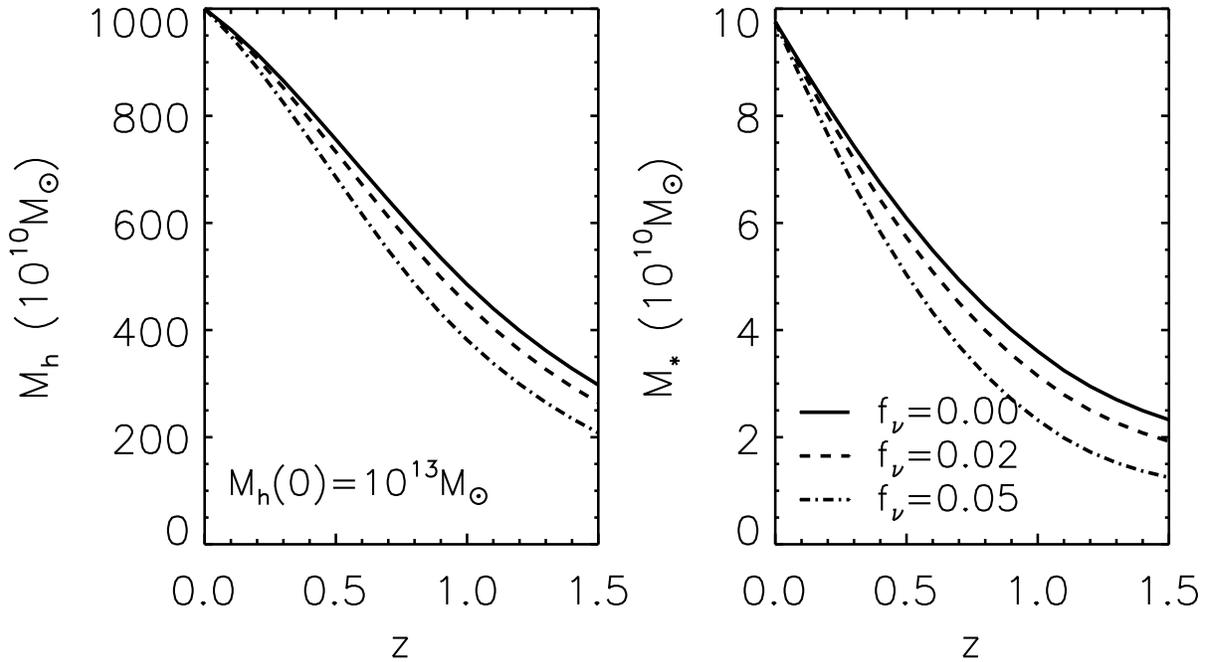}
\caption{Evolution  of the halo mass (left panel) and the stellar mass (right panel) 
accreted from redshift $z$ only through mergers with mass ratio $\xi\ge \xi_{\rm min}=0.03$ 
for three different cases of the neutrino mass fraction: ($f_{\nu}=0,\ 0.02$ and $0.05$ as solid, 
dashed and dot-dashed lines, respectively).}
\label{fig:dMh}
\end{center}
\end{figure}

Since \citet{zhao-etal09} showed that Equations (\ref{eqn:zhao})-(\ref{eqn:zhao3}) 
works on wide mass range at various redshifts for several different cosmological models, 
the applicability of  Equations (\ref{eqn:zhao})-(\ref{eqn:zhao3}) to $\Lambda$MDM cosmology is expected.
Expressing Equation (\ref{eqn:zhao}) in terms of mass and redshift according 
to the chain rule as,
\begin{equation}
\label{eqn:dMhdz_nu}
\frac{dM_{h}}{dz}=
\frac{dM_h}{d{\rm log}\sigma(M_h)} \frac{d{\rm log}\delta_c(z)}{dz}
\frac{d{\rm log}\sigma(M_h)}{d{\rm log}\delta_c(z)}\, ,
\end{equation}
and integrating Equation (\ref{eqn:dMhdz_nu}) over $z$, we finally determine 
the halo mass accretion history $M_{h}(z)$ for a $\Lambda$MDM 
universe. Figure \ref{fig:dMh} shows the evolution of $M_{h}$ and $M_{\star}$ 
via dry mergers with $\xi\ge \xi_{\rm min}=0.03$ for three different cases of $f_{\nu}$ 
in the left and right panels, respectively.  The mass of a descendant halo at present epoch 
is set at $M_{h}(z_{0}=0)=10^{13}\,M_{\odot}$ (corresponding to $M_{\star}\approx 10^{11}\,M_{\odot}$) 
for all three cases. Here, we follow the evolution up to redshift $z\approx 1.5$ since the 
formation epoch of a descendant halo with mass $M_{h}(z_{0}=0)=10^{13}\,M_{\odot}$ is 
approximately $z=1.5$ . As can be seen, both of $M_{h}$ and $M_{\star}$ 
evolve more rapidly for the case of higher value of $f_{\nu}$, and the differences among 
the three cases of $f_{\nu}$ in $M_{h}$ and $M_{\star}$ become larger at higher redshifts.
This result indicates that the high-$z$ progenitor galaxies of a descendant 
halo with the same mass accrete larger amount of DM and stellar masses  
via dry mergers in a $\Lambda$MDM universe than in the $\Lambda$CDM case. 

The physical reason why the halo mass growth rate and the merger rate are higher in a $\Lambda$MDM 
cosmology may be related to the later formation epochs of dark halos \citep[e.g.,][]{SL11}.
The dark halos tend to form later in a $\Lambda$MDM universe than in a $\Lambda$CDM 
universe since the small-scale powers are reduced due to the free streaming effect of massive 
neutrinos. In a $\Lambda$CDM universe the small galactic halos form much earlier than the 
larger halos and thus the interval between the formation epochs of small galactic halos and larger 
halos is long. 
However, in a $\Lambda$MDM universe the formations of dark halos (small and large halos alike) 
are delayed and thus the interval between the formation epochs of small and large halos is 
much shorter  than in the $\Lambda$CDM case. 
In other words, the small halos need to merge faster into the larger halos in a $\Lambda$MDM 
universe.

Now that we have determined the mean merger rate per halo and the mass 
accretion history for the $\Lambda$MDM case, we plug them into 
Equation (\ref{eqn:dMhdzdxi_merg}) to prescribe the first ingredient of our model. 
Regarding the second and third ingredients (i.e., $dR_{e}/dM_{\star}$ and 
$dM_{\star}/dM_{h}$), we use the same prescriptions 
i.e., Equations~ (\ref{eqn:SHMR})-(\ref{eqn:dRedMs}), of the original N12 model 
under the assumption that these relations are still valid for the $\Lambda$MDM case.
It is worth, however, discussing whether or not this assumption is reasonable here.
As mentioned in section \ref{sec:review}, the SHMRs were all determined by 
matching the observational data to the numerical results by applying several different 
techniques such as abundance matching, gravitational lensing and so on, some of which 
are not based on $\Lambda$CDM models. Therefore, even though the SHMRs that 
N12 adopted were obtained  for the $\Lambda$CDM case,  
we expect it to work for the $\Lambda$MDM case.
Regarding  the size-to-stellar mass relation, Equation~(\ref{eqn:dRedMs}), 
it is a local relation obtained by applying the first order galactic dynamics which 
is independent of the background cosmology as long as the gravity is well 
described by the Newtonian dynamics. Therefore, it should be also 
valid for the $\Lambda$MDM case.

\begin{figure}
\begin{center}
\plotone{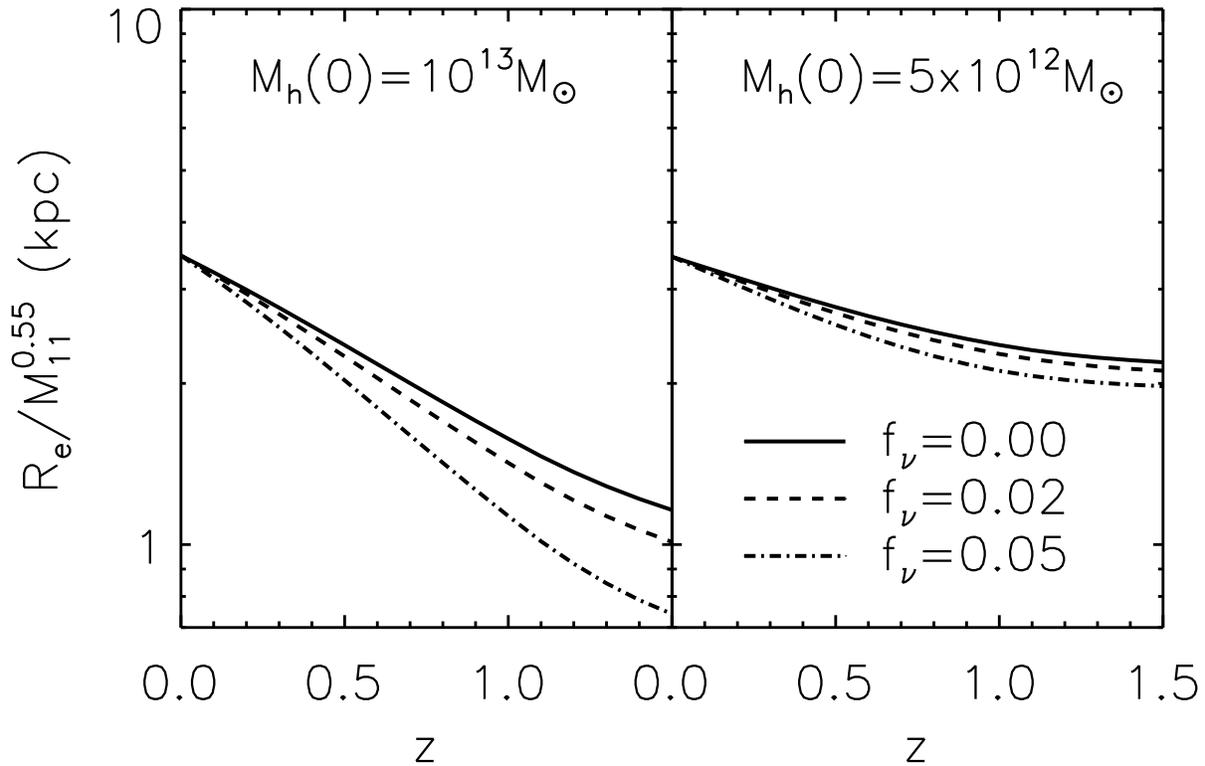}
\caption{Compactness evolution of the progenitor galaxies for the case of the massive descendant halo  
with $M_{h}(0)=10^{13}\,M_{\odot}$ (left panel) and  for the case of the less massive descendant 
halo with $M_{h}(0)=5\times 10^{12}\,M_{\odot}$ (right panel). In each panel, the solid, dashed 
and dot-dashed lines correspond to $f_{\nu}=0,\ 0.02$ and $0.05$, respectively.}
\label{fig:Re}
\end{center}
\end{figure}

Plugging the prescribed ingredients into Equations (\ref{eqn:d2Msdzdxi})-(\ref{eqn:d2Redzdxi}),
we finally calculate the size and stellar mass evolution of the progenitor galaxies of a
descendant halo observed at present epoch in a $\Lambda$MDM universe and then 
determine the compactness of the progenitor galaxies at each redshift as $R_{e}M^{-0.55}_{11}$.
Figure \ref{fig:Re} shows how the compactness of the progenitor galaxies of 
a descendant halo observed at $z=0$ evolves in a $\Lambda$MDM universe. The left and right 
panel corresponds to the cases of $M_{h}(0)=10^{13}\,h^{-1}M_{\odot}$ and 
$M_{h}(0)=5\times 10^{12}\,h^{-1}M_{\odot}$, respectively. 
In each panel, the solid, dashed and dot-dashed lines correspond to the cases 
of $f_{\nu}=0,\ 0.02$ and $0.05$, respectively. 
As can be seen, the high-$z$ progenitors become more compact for the 
higher-$f_{\nu}$ case, which confirms that the presence of massive neutrinos 
plays a role of promoting the size growth of the massive ETGs at high 
redshifts. 
Furthermore, this effect of massive neutrinos on the size evolution
is stronger for the case that $M_{h}(0)$ has a higher value, which is consistent with the 
observational result that the most compact high-$z$ ETGs are usually the most 
massive ones \citep{cimatti-etal12,daddi-etal05,nipoti-etal12,trujillo-etal06,
vandokkum-etal08}.

\begin{figure}
\begin{center}
\plotone{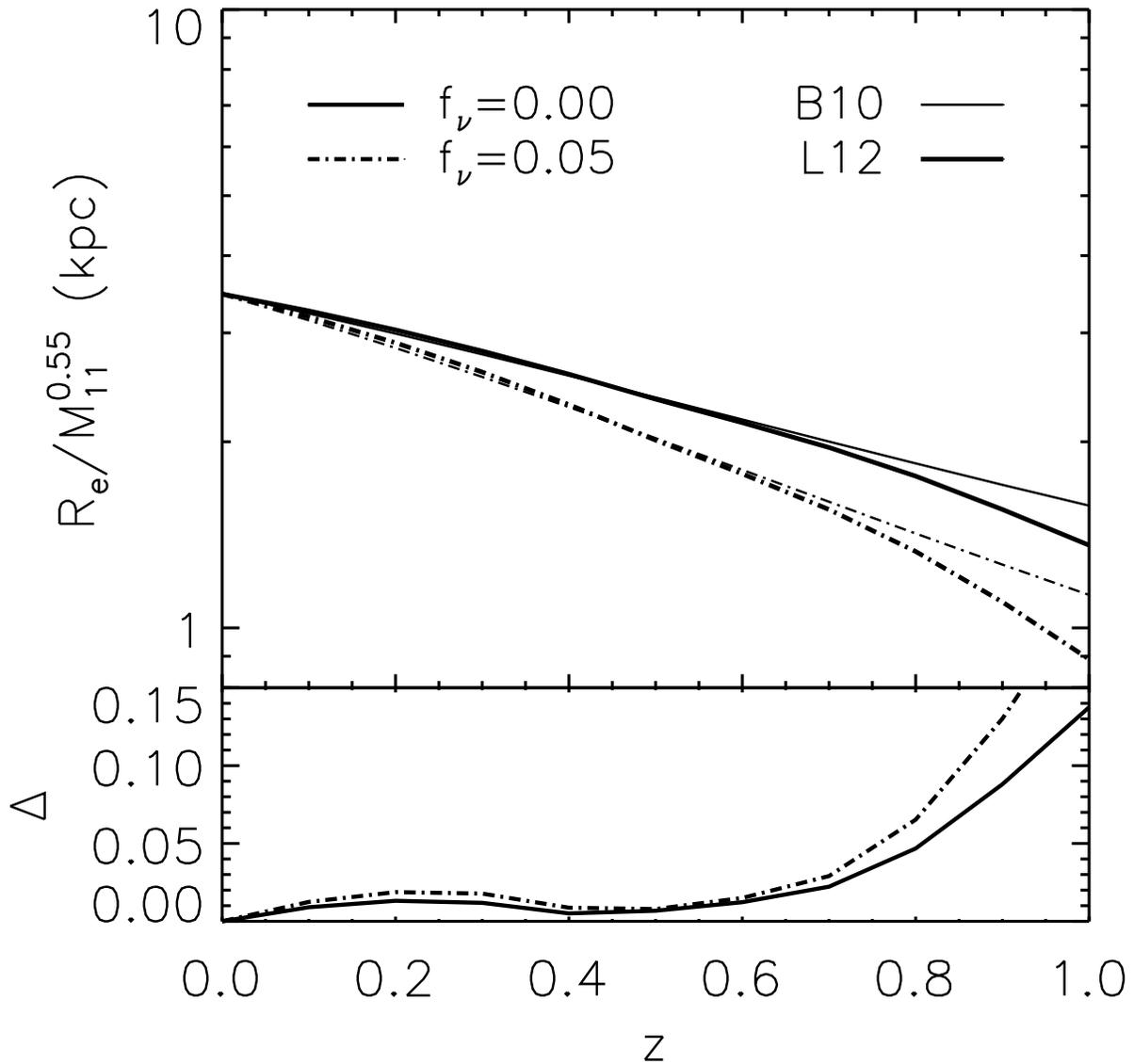}
\caption{(Top panel): Comparison between the compactness evolution results obtained by using 
two different SHMRs: the case of using B10 SHMR as thin lines while the case of using L12 SHMR as 
thick lines. For each case, the solid and dashed lines correspond to $f_{\nu}=0$ and 
$f_{\nu}=0.05$, respectively. (Bottom panel): Fractional difference between the thin and thick curves 
shown in the top panel for the two cases of $f_{\nu}$.}
\label{fig:nRe}
\end{center}
\end{figure}

The results shown in Figure \ref{fig:Re} have been obtained by using the B10 
SHMR. To examine whether or not the use of a different SHMR will change 
the trend, we repeat the calculation of the compactness evolution but with 
using the L12 SHMR.  The top panel of Figure \ref{fig:nRe} shows the 
compactness evolution obtained by using the L12 SHMR as thick solid and 
dashed lines for the cases of $f_{\nu}=0$ and $f_{\nu}=0.05$, respectively. The results 
shown in the left panel of Figure \ref{fig:Re} (only for the two cases of 
$f_{\nu}=0$ and $0.05$) are also plotted as thin lines for comparison.  Since 
the L12 SHMR was originally obtained for the galaxies at $z\le 1$, we show 
the results only at $z\le 1$. The bottom panel of Figure \ref{fig:nRe} shows the fractional 
difference between the two SHMR cases. 
As one can see, the use of a different SHMR does not destroy the trend that the 
progenitor compactness evolves faster for a higher $f_{\nu}$ case, which 
indicates that our result is qualitatively robust against the uncertainties of SHMR. 
However, it has to be also noted that using a different SHMR yields quantitatively 
a different compactness evolution of the progenitors and thus it will be important 
to refine the SHMR as accurately as possible.

We have so far studied theoretically the progenitor evolution  of one single descendant halo existent 
at $z=0$ .  The observed ETGs shown in \citet{cimatti-etal12} must correspond not to the 
progenitors of one single descendant halo existent at $z=0$ but to the progenitors of different 
descendants existent at different redshifts.  Therefore, to make a comparison with the observational 
results, it is necessary to model theoretically the progenitor evolutions of different descendant 
halos existent at different redshifts for each case of $f_{\nu}$.  
We consider only those progenitors whose distribution in the $M_{\star}$-$z$ plane are similar to 
the observed ETGs shown in Figure of \citet{cimatti-etal12}, having stellar masses in the range 
of $10^{10.5}\le M_{\star}/M_{\odot}\le 10^{12}$ at redshifts of $0\le z\le2.5$.  It amounts to considering 
those halos as representative descendants whose total mass lie between $5\times 10^{12}\,M_{\odot}$ 
and $2\times 10^{13}\,M_{\odot}$. 

\begin{figure}
\begin{center}
\plotone{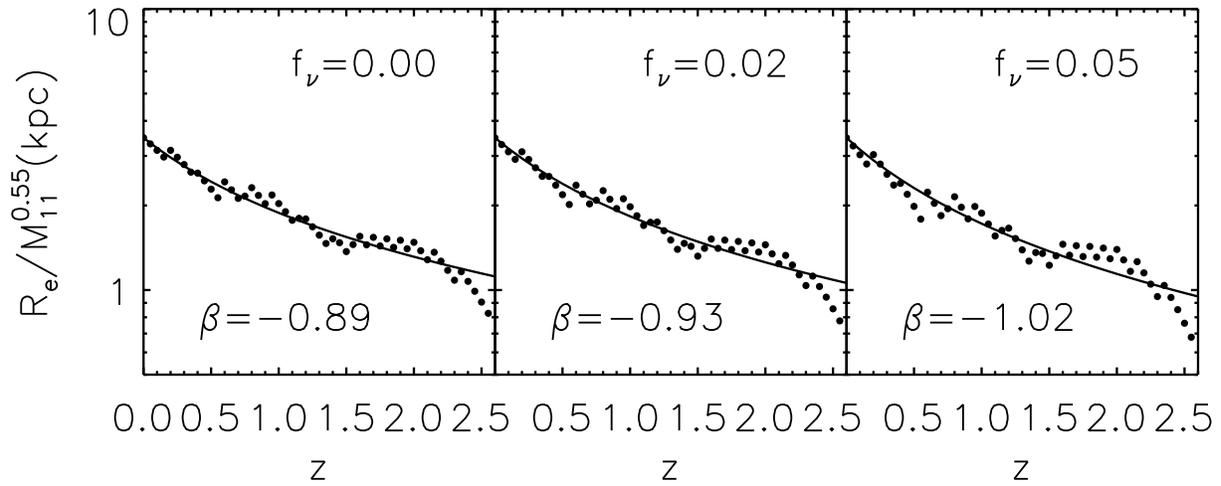}
\caption{Average compactness of the progenitor galaxies (dots) with stellar masses 
in the range of $10^{10.5}\le M_{\star}/M_{\odot}\le 10^{12}$ for three different cases of $f_{\nu}$. 
The average is taken over the progenitor histories of representative descendant halos whose 
progenitors have the same stellar mass distributions in the redshift range of $0\le z\le 2.5$ as 
the observational results shown in Figure 2 of \citet{cimatti-etal12}.  
In each panel, the solid line corresponds to the best-fit scaling relation of 
$R_{e}/M_{11}^{0.55}=(1+z)^{\beta}$. }
\label{fig:beta}
\end{center}
\end{figure}

Approximating the compactness evolution of the theoretically modeled progenitors as the scaling relation of 
$R_{e}M_{11}^{-0.55}=(1+z)^{\beta}$, we determine the best-fit value of $\beta$ with the help of the $\chi^{2}$ 
minimization method for each case of $f_{\nu}$.  Figure \ref{fig:beta} shows the best-fit scaling relations 
(solid line) and compare them with the average compactness evolution of the modeled progenitor galaxies (dots) 
for three different cases of $f_{\nu}$. As can be seen, the absolute value of $\beta$ increases from $0.89$ to 
$1.02$ as $f_{\nu}$ increases from $0$ to $0.05$.    
According to \citet{cimatti-etal12}, the absolute value of $\beta$ obtained from the 
observed ETGs at redshifts of $0\le z\le 2.5$ is close to unity. At face value it suggests that the 
observational result is consistent with the theoretically estimated value of $\beta$ for the 
$\Lambda$MDM model with $f_{\nu}=0.05$. 

\section{DISCUSSION AND CONCLUSION}
\label{sec:con}

We have shown that the presence of massive neutrinos plays a role of enhancing 
the mean merging rate per halo as well as the merger-driven increment 
in halo mass on the massive galaxy scale and thus that the ETG compactness 
evolves much more rapidly in a $\Lambda$MDM universe.
It is worth discussing here whether the $\Lambda$MDM prediction of the higher 
merging rate per halo is consistent with the observational indication. 
N12 mentioned that since the inferred merging rate from the observations 
of binary galaxy systems is even lower than the $\Lambda$CDM prediction, 
enhancing the merging rate should not be the key to explaining away the 
anomalous compactness of the high-$z$ ETGs. However, very recently, 
\citet{jian-etal12} have demonstrated that the merging rate inferred 
from the observed galaxy pairs might be severely contaminated by the 
spurious projection effect.

However, our current model is only a reasonable approximation based on 
several simplified assumptions to the true effect of massive neutrinos on 
the ETG evolution. It will be definitely necessary to improve and refine our model 
by incorporating more realistic assumptions and adopting more accurate 
prescriptions. 
In the first place, it will be essential to determine more accurately the mean 
merger rate per halo by using high-resolution N-body simulations for a 
$\Lambda$MDM universe. Although the ellipsoidal EPS model that we have used 
to calculate the mean merger rate per halo is a significantly improved 
version of the original spherical EPS model, it still suffers from maximum 
$20\%$ errors when compared with the N-body results \citep{zhang-etal08}. 
The size-to-stellar mass relation also has to be improved by accounting for 
the following realistic aspects of the true merging process of progenitor 
galaxies: (i) the occurrence of dissipational wet mergers; (ii) the time 
lapse between the moment of halo merging and that of galaxy merging; and 
(iii) the presence of disc shaped progenitors. High-resolution gas 
simulations with massive neutrinos included will be required to address 
these complicated issues. 

Another thing that may deserve discussing is the question of whether or 
not the $\Lambda$MDM model is the only solution to the anomalous strong 
size growth of the high-$z$ massive ETGs.  
One might think that different cosmologies such as models with primordial non-Gaussianity, 
dynamic dark energy or modified gravity might also affect the mean merger rate per halo, 
thus influencing size growth of the massive high-z ETGs.
To examine if any other cosmological models could be also a solution,
however, it will require much more works, 
totally renewing all prescriptions  since in these models 
the extended EPS formalism is no longer valid, 
the gravitational dynamics changes and the SHMR could be also quite different.  

Our final conclusion is that the tension of the standard structure formation scenario 
with the observed anomalous compactness of the massive high-$z$ ETGs can be 
greatly alleviated without changing the key cosmological parameters from the WMAP7 
values if the presence and effect of massive neutrinos are properly taken into account.  

\acknowledgments

We thank a referee whose useful comments helped us improve the original manuscript.
This work was supported by the National Research Foundation of Korea (NRF) 
grant funded by the Korea government (MEST, No.2012-0004195). 
Support for this work was also provided by the National Research Foundation of Korea to the
Center for Galaxy Evolution Research. (NO. 2010-0027910)

\newpage
\clearpage
\clearpage
\clearpage
\clearpage

\end{document}